\documentclass[a4paper, 11pt]{article}
\usepackage[a4paper, total={6in, 8in}]{geometry}
\usepackage[utf8]{inputenc}
\usepackage[T1]{fontenc}
\usepackage{graphicx,epstopdf} 
\usepackage{hhline}
\usepackage{float}
\usepackage{subcaption}
\usepackage{titlesec}
\usepackage{tikzpagenodes}
\usepackage{lipsum}
\usepackage{titling}
\usepackage{algorithm}
\usepackage[noend]{algorithmic}
\usepackage{eqparbox}
\usepackage{sidecap}
\usepackage{comment}
\usepackage{bm}
\usepackage{booktabs}
\usepackage{wrapfig}
\usepackage{multirow}

\usepackage{xcolor}
\usepackage{amsmath}
\usepackage{textcomp, gensymb}
\usepackage{gensymb}
\usepackage{amssymb}
\usepackage{mathtools}

\usepackage[export]{adjustbox}
\usepackage{lipsum}
\usepackage{pgf}
\usepackage{rotating,multirow}
\usepackage{float} 
\usepackage{authblk}
\usepackage{xcite}
\usepackage{xr}

\title{Impacts of poroelastic spheres}

\author[1,2]{Benjamin Gorin}
\author[3]{Neil Ribe}
\author[2]{Daniel Bonn}
\author[1]{Hamid Kellay}

\affil[1]{Université de Bordeaux, LOMA, UMR CNRS 5798, Talence, 33405, France}
\affil[2]{University of Amsterdam, Van der Waals Zeeman institute, Amsterdam, 94485, Netherlands}
\affil[3]{Université Paris Saclay, FAST, UMR CNRS 7608, Orsay, 91405, France}
\date{}

\begin{document}
\maketitle
\begin{abstract}

We study experimentally the impact on rigid surfaces of different soft porous solids saturated with liquid: hydrogel balls and liquid-saturated foam balls. 
The static contact of such soft solids with the substrate is well described by Hertz contact theory. However, their rebound behavior can only be explained by invoking a variety of dissipation mechanisms. We find that the restitution coefficient of soft porous balls generally increases with the impact velocity. We propose that this behavior can be explained by a combination of three `wet' dissipation mechanisms: capillary adhesion, viscous dissipation in a liquid film between the ball and the substrate, and poroelastic dissipation due to porous flow inside the ball. While the first two dissipations are known, we present a new theory for poroelastic dissipation, and show that it allows experimental data for saturated foam balls to be reduced to a master curve against a suitably normalized impact velocity.
The understanding of this dissipation mechanism with its dependence on both permeability of the porous solid and liquid viscosity can open the way towards engineering a new generation of shock absorbers and cushions.  

\end{abstract}

\thispagestyle{empty}

\section*{Introduction}


In an elastic collision, kinetic energy is transformed into stored elastic energy, which is then transformed to kinetic energy again. However, collisions are imperfectly elastic and energy dissipation due to different mechanisms occurs. 
Here we study such energy dissipation in bouncing spheres: our objective is to unravel the intricate interplay of different energy dissipation forms in the particular case of wet poroelastic spheres, specifically liquid filled foam balls and hydrogel spheres. Bouncing of hydrogel balls have been first studied in 2003 by Tanaka \cite{tanaka_bouncing_2003,tanaka_impact_2005} who revealed that hydrogel spheres can undergo strong deformations upon impact without cracking and measured a strong variation of the coefficient of restitution with the impact velocity: Part of the initial kinetic energy is dissipated during the deformation. He also measured the reaction force of a deformed hydrogel sphere and showed that it deviates from Hertz's law for strong deformations. Later, Waitukaitis et.al \cite{waitukaitis_bouncing_2018} studied bouncing of hydrogel spheres upon a hot surface and measured restitution coefficients which can be higher than unity. They explained these measurements by the coupling between the elastic deformation and the Leidenfrost effect coming from the evaporation of the water film covering the hydrogel. As the hydrogel is filled with water and is permeable, the evaporated water film is regenerated between each bounce and remains present for multiple rebounds above the hot surface.
Conversely, impacts of filled open-cell foam balls have not yet been studied to our knowledge. However, deformations of foam materials have been studied previously under different conditions due to their applications in shock absorbers. In 1966, Gent and Rush \cite{gent_viscoelastic_foam_1966} first proposed a theoretical description of the energy loss of a filled foam material under mechanical oscillatory deformation. They revealed that the liquid flow plays a major role in the energy loss as it is being forced to go in and out of the foam during a cyclic deformation. Later, Hilyard \cite{hilyard_observations_1971} studied impacts on a liquid filled foam material and in particular the contribution of the fluid flow on the impact behaviour. As the impact velocity increases, Hilyard observed an increase followed by a decrease of the ratio between the dissipated and stored energy within the elastic liquid filled foam.
More recently, another study of the contribution of the fluid flow in porous material has been proposed by Dawson et.al \cite{dawson_dynamic_2008}. In this experimental work, they also studied the contribution of the fluid to the elastic response of a liquid filled foam cylinder under slow compressive strain rate. They showed that in the elastic regime, the liquid flows through the porous material and induces an additional pressure which follows Darcy's law.
However, the energy dissipation due to the internal flow of the viscous fluid within the porous medium has not yet been investigated.
In the experiments we report here, we study impacts of fluid filled poroelastic spheres and find remarkable agreement with Hertz contact theory for small deformations of the sphere upon impact on a solid surface: the elastic energy stored during the impact is not affected by the presence of the internal fluid. Still, the restitution coefficient strongly depends on the impact velocity. This allows to use a simple approach in which porous flow within the sphere is driven by the gradients in elastic pressure predicted by Hertz's theory allowing to calculate analytically the effect of the poro-elasticity on the restitution coefficient of the spheres impacting solid substrates. This theory is then compared to experiments and very good agreement is found.

In 1881, Heinrich Hertz published his pioneering work \cite{hertz_ueber_1882,landau_theory_1986} on the contact of two elastic ellipsoids. He gave an analytical expression for the force on a deformed elastic sphere in contact with a rigid surface. 
In elastic collisions, the initial kinetic energy ($ E_{k,0} \sim m{V_0}^2$) is entirely restored as kinetic energy in the rebound ($E_{k,r} \sim m{V_R}^2$). This is so if there is no dissipation of energy during the impact, giving rise to a restitution coefficient $e=1$. The coefficient of restitution $e$ is defined as the ratio between the rebound and the impact velocities $V_R$ and  $V_0$ respectively.
We preferably focus on the square of the coefficient of restitution which corresponds to the relative kinetic energy left after the impact.
\begin{equation}
    e^2=\frac{{V_{R}}^2}{{V_{0}}^2}   
\end{equation}

In reality, collisions are inelastic and a fraction of the incident energy is dissipated ($E_{k,r}=E_{k,0}-E_{d}$) during impact \cite{lifshitz_experiments_1964,montaine_coefficient_2011}, so that $e^2<1$. The kinetic energy of a dry ball impacting a surface can be dissipated through different mechanisms including plastic deformation \cite{lifshitz_experiments_1964,johnson_mechanics_1998,thornton_theoretical_1998,thornton_elastic-plastic_2017}, frictional contacts due to the surface roughness \cite{montaine_coefficient_2011}, propagation of elastic waves \cite{hunter_energy_1957,reed_energy_1985} and adhesive forces \cite{dahneke_capture_1971}. 
The ratio between the energy dissipated from the solid phase and the kinetic energy before the impact, denoted $\alpha_{dry}=\frac{E_{d,dry}}{E_{k,0}}$, is not investigated here but is taken into account by defining $e_{ref}$ as follows :
\begin{equation}
    {e_{ref}}^2=1-\alpha_{dry}
\end{equation}

When a liquid film covers the surface of the ball, additionnal dissipative effects are involved in the bouncing behaviour.
 
In 1997, Johnson, Kendall and Roberts \cite{johnson_surface_1997} modified the Hertz theory by incorporating the effect of adhesion. This is known as the JKR theory and is of interest for understanding aggregation problems or pull-off forces between two bodies. When the substrate and/or the ball's surface is wet, the presence of a liquid film between the ball and the substrate affects the rebound of the ball as energy is required to overcome capillary forces during separation of the ball from the substrate. An expression for the role of the capillary adhesion force between two spheres has been proposed by Gollwitzer
et.al \cite{gollwitzer2012coefficient}.
This expression involves an adhesion force proportional to cos($\theta$) with $\theta$
being the contact angle of the liquid bridge in contact with the surface. During the
approach phase, we can assume that the contact angle is around 90° as the liquid film
spreads over the surface so the capillary force mainly plays a role during the
rebound phase. Therefore, the energy cost needed to separate the ball from the surface is equivalent to the energy needed to create two new liquid-air interfaces is $E_{\gamma}\sim \gamma {D_{max}}^2$ \cite{johnson_surface_1997}, where $\gamma$ is the surface tension of the liquid-air interface and $D_{max}$ is the maximum contact diameter of the ball with the substrate.
If we assume that adhesion forces only affect the rebound phase and that $D_{max}$ can be estimated using the Hertz model, the coefficient of restitution $e$ is found from the equation of energy conservation after taking into account the energetic contribution of the capillary adhesion:
\begin{equation}
      e^2 \sim {e_{ref}}^2-\frac{4\gamma \pi R\left(m/k\right)^{2/5}}{m}V_0^{-6/5}
      \label{eq: COR adhesion model - intro}
\end{equation}

In equation (\ref{eq: COR adhesion model - intro}), $R$ and $m$ are respectively the radius and the mass of the ball and $k={16 \sqrt{R}E_0}/{15 \left(1-\nu^2 \right)}$ is a constant that depends on the elastic modulus $E_0$ and the Poisson ratio $\nu$ of the ball. For convenience we rewrite equation (\ref{eq: COR adhesion model - intro}) as:

\begin{equation}
     e^2 = {e_{ref}}^2 - \alpha_{Adh}
     \label{eq:1-alpha1}
\end{equation}

Below we shall call equation (\ref{eq:1-alpha1}) the `capillary adhesion' dissipation model. Details of the model are given in Supplementary Information.

During the contact of the wet ball with the surface, the thin liquid film formed between the surface and the ball is squeezed resulting in additional viscous losses:
This situation is known as elastohydrodynamic collision \cite{davis_elastohydrodynamic_1986,barnocky_elastohydrodynamic_1988,gondret_experiments_1999,gondret_bouncing_2002,davis_elastohydrodynamic_2002}. Based on lubrication theory, Davis et.al \cite{davis_elastohydrodynamic_2002} proposed a simple model to account for the dissipation in the viscous film. They showed that the rebound of the sphere depends on the Stokes number $St=m V_0/(6 \pi \eta R^2)$ where $\eta$ is the viscosity of the liquid film and that no rebound occurs below a critical Stokes number $St_c$. They found that $e$ is given by
\begin{equation}
    e^2 = \left( 1-\frac{St_c}{St}\right)^2 
    \label{eq: COR EHL model - intro}
\end{equation}

The critical Stokes number $St_c$ is defined as $St_c=\frac{x_0}{x_r}$ with $x_0$ the initial film thickness and $x_r$ the minimum distance between the solid sphere and the surface during the contact which depends on the liquid film properties (thickness, viscosity) and the elastic properties of the ball \cite{davis_elastohydrodynamic_1986}.
Details of the model are given in Supplementary Information. We
shall henceforth call equation (\ref{eq: COR EHL model - intro}) the `viscous film' dissipation model. 
For convenience, we rewrite it as:
 \begin{equation}
      e^2 = {e_{ref}}^2 - \alpha_{Film}
      \label{eq:1-alpha2}
 \end{equation}
The total effect of capillary adhesion plus viscous film dissipation is obtained by combining equations (\ref{eq:1-alpha1}) and (\ref{eq:1-alpha2}).
\begin{equation}
    e^2 = {e_{ref}}^2 - \alpha_{Adh} - \alpha_{Film}
    \label{eq:1-alpha1-alpha2}
\end{equation}

Impacts of wet spheres have been studied previously \cite{davis_elastohydrodynamic_1986,johnson_surface_1997,davis_elastohydrodynamic_2002,gollwitzer2012coefficient,cruger2016coefficient,vallone2023dynamics} and different mechanisms have been proposed to explain the bouncing behaviour of such spheres. In addition to previously well-studied dissipation mechanisms such as capillary adhesion \cite{johnson_mechanics_1998,vallone2023dynamics} and thin-film viscous dissipation \cite{davis_elastohydrodynamic_1986,davis_elastohydrodynamic_2002}, we propose a new model for the dissipation due to the coupling between the porous ball's deformation and internal flow within the poro-elastic material.

The third dissipation model we shall consider is a `poroelastic' model. 
If the ball is porous and fluid-filled, its deformation during impact will induce pressure gradients that drive internal porous flow, producing additional viscous dissipation \cite{terzaghi25, biot41japplphys,cryer63qjmam}.
The derivation of this model is given below in the Results section. As we shall show, poroelastic dissipation is important and even dominant in some cases.

\section*{Results}

Our experiments make use of different balls, some of which are porous and can be filled with liquids of different viscosities. As a dry reference, we use so-called `bouncy balls', commercial rubber balls with restitution coefficients close to unity. We use two different types of wet balls. The first are hydrogel spheres, which are composed of a water-swollen polymer network and which also have high restitution coefficients. Hydrogels are materials which have generated much recent interest for applications ranging from agriculture (water storage in the soil) to the biomedical field \cite{osada_soft_1998,gong_double-network_2003},  where materials that can undergo strong deformations without cracking are needed \cite{haque_super_2012}. Second, we use commercial foam balls made of polyurethane. Such balls deform readily upon impact, making them good candidates for investigating impacts of soft porous balls. Compared to hydrogels, our foam balls are larger and sponge-like, allowing them to be filled with liquids of different viscosities.
The solid surface used is a 1.5 cm thick transparent poly-carbonate plate. 

To quantify the elastic deformation of the three types of balls during impact, we measured the contact diameter as a function of time using a fast camera placed below the transparent solid surface (see Methods). To compare these data with the predictions of the Hertz model, we need to know the effective elastic modulus $E^*= E_0/(1-\nu^2)$ where $E_0$ is Young's modulus and $\nu$ is Poisson's ratio. We measured this modulus by placing the balls under compression in a rheometer (Methods). We found $E^* = 2$ MPa for the rubber ball, 45 kPa for the hydrogel, and 130 kPa for the foam balls independently of the filling liquid. In the Methods section we demonstrate that
the elastic deformation during impact of all three types of balls is well described by the Hertz contact model. 
Note that in our experiments, we use relatively low impact velocities so that the deformation of the balls are small and can be described by the Hertz model. For higher impact velocities, the deformation of the soft hydrogel balls or the foam balls becomes too important and the Hertz contact theory breaks down \cite{tanaka_bouncing_2003,tanaka_impact_2005}. In this large deformation regime, the coefficient of restitution shows a strong decrease for both the hydrogel and the foam balls (see Figure \ref{fig:COR highspeed} in Supplementary Information); this effect has been previously studied for hydrogel spheres \cite{tanaka_bouncing_2003,tanaka_impact_2005}.

\begin{figure}[ht]
    \centering
    \includegraphics[width=1\textwidth]{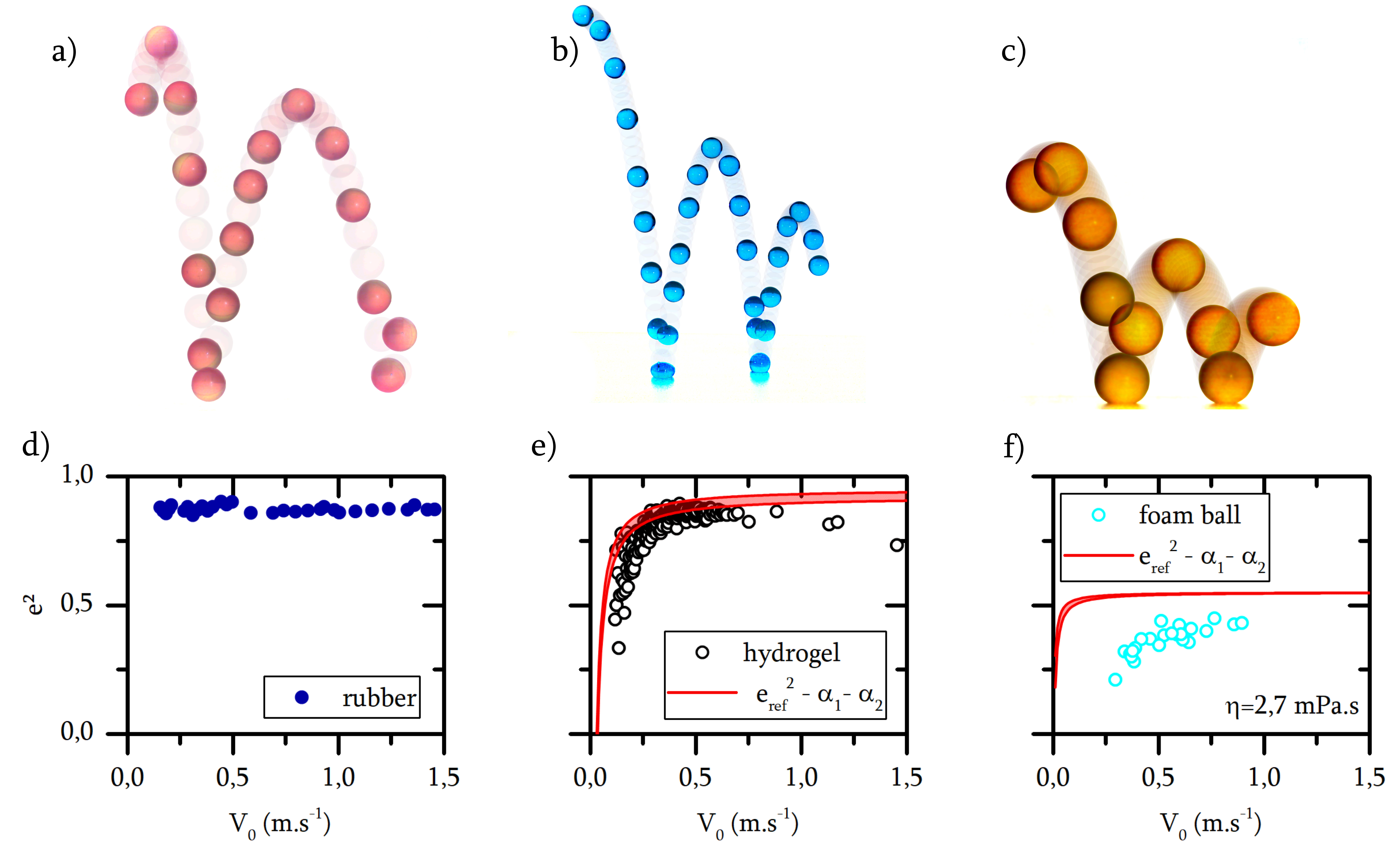}
    \caption{Bouncing of (a) a rubber ball ($R=21$ mm, frame rate 50 fps), (b) a hydrogel ball ($R=7$ mm, 280 fps) and (c) a dry foam ball ($R=34$ mm, 500 fps). Each panel is a superposition of images recorded with a fast camera. Some of these images are highlighted to illustrate the trajectory. (d) Coefficient of restitution of the rubber ball. e) Coefficient of restitution of the hydrogel ball (black circles) compared with the model of equation (\ref{eq:1-alpha1-alpha2}) with $6 < St_c < 10$ and $0.92< e_{ref}<0.97$. (f) Coefficient of restitution of a foam ball filled with water-glycerol mixture I (liquid properties are given in table \ref{Table:  liquids_properties} in Methods), compared with the model of equation (\ref{eq:1-alpha1-alpha2}) with $e_{ref}=0.55$ corresponding to the dry case and $6 <St_c <10$.}
    \label{fig:bouncing-overview}
\end{figure}

In Figure \ref{fig:bouncing-overview}, the rebound of our three different balls (rubber, hydrogel, and foam) is illustrated as a series of superimposed snapshots. 
The rubber ball rebounds almost to its initial height, the fluid-filled foam ball rebounds to only about half its initial height after the first bounce, and the hydrogel ball is intermediate.
    
Figure \ref{fig:bouncing-overview}  (d)-(f) show how the coefficient of restitution $e^2$ varies as a function of the impact velocity $V_0$ for our three types of balls. For rubber balls, $e^2$ is independent of $V_0$ (Figure \ref{fig:bouncing-overview}d). However, $e^2$ varies strongly as a function of $V_0$ for the other two types of ball. For hydrogel balls, $e^2$ increases rapidly with increasing $V_0$ up to 0.9 m s$^{-1}$, and then decreases slowly thereafter (Figure \ref{fig:bouncing-overview}e). For the foam balls saturated with a water-glycerol mixture, $e^2$ is a monotonically increasing function of $V_0$ (Figure \ref{fig:bouncing-overview}e) in the range of velocities used in this study. For higher impact velocities, the assumption of the Hertz model of small deformations of the sphere is no longer valid \cite{tatara_compression_1991,tanaka_impact_2005}. In this study, we only consider small impact velocities: the indentation of the sphere $h$ remains small compared to the radius of the ball $h/R << 1$ so that the bouncing behaviour is well captured by the Hertz model.

It may seem counter-intuitive that increasing the impact speed of the ball results in higher restitution. The coefficient of restitution is defined as the ratio $e^2=\frac{E_k-E_{d}}{E_k}$. The absolute value of energy dissipated from capillary adhesion and viscous dissipation in the liquid film $E_d$ increases with the velocity but the relative dissipated energy $\frac{E_d}{E_{k,0}}$ decreases with the impact velocity. The observed variation of the coefficient of restitution for the hydrogel balls can be explained reasonably well by a combination of the capillary adhesion and viscous film dissipation models.
The resulting prediction of $e^2$ is indicated by the narrow red regions in Figures \ref{fig:bouncing-overview} e).
Because $e_{ref}$ and $St_c$ are not known for the hydrogel spheres, we used ranges of values $0.92 < e_{ref} < 0.975$ and $6 < St_c < 10$. The interval of $St_c$ is chosen based on the assumptions of the maximum and minimum liquid film thickness. The maximum liquid thickness is chosen as $x_0=1$ mm and the minimum thickness is given by the roughness of the surface which is of the order of $10^{-6}$ m \cite{range_influence_1998}. Although we only consider two mechanisms of energy dissipation, the result of equation (\ref{eq:1-alpha1-alpha2}) gives the good trend of the variation of $e^2$ for a large range of velocities.

For the saturated foam ball, the combination of capillary adhesion and viscous film dissipation does not explain the observed restitution coefficients, assuming $e_{ref}$ to  be that of a dry foam ball (see below) and $6 < St_c < 10$. We propose that our observations for foam balls require a third dissipation mechanism associated with porous flow inside the fluid-saturated ball, which has much larger pores than the hydrogel spheres. In the next subsection we present a new model for this poroelastic dissipation mechanism.  
 
\subsection*{Poroelastic dissipation}

The theory of poroelasticity has a long history going back to the work of K. Terzaghi \cite{terzaghi25} and M. Biot \cite{biot41japplphys}. In this theory, the deformation of the solid elastic matrix is influenced by gradients of interstitial fluid pressure, and the fluid pressure itself evolves with time. The physics of this situation gives rise to a complicated initial/boundary-value problem that would have to be solved numerically. 
However, in the experiments we report here on deforming poroelastic spheres, we find remarkable agreement with the loading curves and maximum contact diameter predicted by Hertz contact theory. This suggests that we try a simpler approach in which porous flow within the sphere is driven by the gradients in elastic pressure predicted by Hertz's theory.

The flow in a porous medium is described by Darcy's law,
\begin{equation}
    u=\frac{\kappa}{\phi \eta}\nabla P
    \label{eq: Darcy's law}
\end{equation}
where $u$ is the pore-scale percolation velocity, $\kappa$ is the permeability (m$^2$), $\eta$ is the fluid viscosity (Pa.s), $\phi$ is the porosity and $\nabla P$ is the hydraulic pressure gradient (Pa.m$^{-1}$). The effective flux of fluid through the medium (volume of fluid per unit surface area per unit time) is the product $\phi u$. During impact, deformation of the ball around the contact area gives rise to a pressure gradient that drives porous flow. The pressure scales as the Hertzian contact force $F_{Hertz}$ divided by the contact area of radius $r$, or $P \sim F_{Hertz}/r^2$. Let $\delta$ be the height of portion of the ball above the contact area in which the pressure gradient is applied (see Figure \ref{fig5: poroelastic model}). 
We can then rewrite equation (\ref{eq: Darcy's law}) as

\begin{equation}
    u=\frac{\kappa}{\phi \eta}\nabla P \sim \frac{\kappa}{\phi \eta} \frac{F_{Hertz}}{r^2 \delta}
    \label{eq: Darcy modified}
\end{equation}
From Hertz contact theory, we have
\begin{equation}
    F_{Hertz} = \sqrt{\frac{Rh^3}{D^2}} \quad \quad and \quad \quad r=\sqrt{Rh}
    \label{e: Hertz equations}
\end{equation}
where $D=3 \left( 1-\nu^2\right)/4E$, $E$ is Young's modulus, $h$ the deformation of the poroelastic ball and $\nu$ is the Poisson's ratio.
Replacing in equation (\ref{eq: Darcy modified}) the Hertz force and the contact radius and assuming $\delta\sim h$, we have
\begin{equation}
    u \sim \frac{\kappa}{\phi \eta} \sqrt{\frac{1}{D^2 R h}}
\end{equation}
From Bercovici et.al \cite{bercovici_energetics_2003}, the rate of viscous dissipation per unit volume is :
\begin{equation}
    \psi = \frac{\eta \phi^2}{\kappa}u^2\sim \frac{\kappa}{\eta D^2 R h}
    \label{eq:rate of viscous dissip}
\end{equation}
The volume $\Omega$ over which the dissipation occurs scales as
\begin{equation}
    \Omega \sim r^2 h \sim R h^2
    \label{eq: volume dissip}
\end{equation}
From equations (\ref{eq:rate of viscous dissip}) and (\ref{eq: volume dissip}), the total rate of viscous dissipation inside the sphere is:

\begin{equation}
    \Phi \sim \psi R h^2 \sim \frac{ \kappa h}{\eta D^2}
    \label{eq: total rate of viscous dissip}
\end{equation}
The total dissipation rate over the contact time $t_c$ (which includes both the compression and rebound phases) is
\begin{equation}
\Delta E \sim \frac{\kappa}{\eta D^2} \int_0^{t_c} h\mathrm d t.   
\end{equation}
$\Delta E$ can be estimated as
\begin{equation}
\Delta E\sim \frac{\kappa}{\eta D^2} h_{max} t_c \sim \frac{\kappa}{\eta D^2}\left(\frac{m V_0^2}{k}\right)^{2/5}
\left(\frac{m^2}{k^2 V_0}\right)^{1/5}
 = B \frac{\kappa}{\eta D^2}\left(\frac{m}{k}\right)^{4/5} V_0^{3/5}
\end{equation}
where we have used the scales for $h_{max}$ and $t_c$ from Hertz contact theory \cite{landau_theory_1986} and $k = 4\sqrt{R}/5D$. The quantity $B$ is the unknown constant of proportionality.

For convenience and later use, we define a characteristic velocity $V_c$ as

\begin{equation}
    V_c= \left( \frac{\kappa^5}{\eta^5D^{10}mk^{4}} \right)^{1/7}
    \label{eq: Vc}
\end{equation}

The coefficient of restitution is then 
\begin{equation}
e^2 = {e_{ref}}^2 - \frac{\Delta E}{(1/2) m V_0^2}\equiv {e_{ref}}^2 - \alpha_{Poro}.
\label{eq: COR poroelastic model}
\end{equation}

Note that $\alpha_{Poro} \propto V_0^{-7/5}$, which is consistent with an increase of the coefficient of restitution with increasing $V_0$. 
Then, we write the full expression of the coefficient of restitution of wet poroelastic balls which includes the three different dissipative mechanisms (capillary adhesion, viscous dissipation in the liquid film, viscous dissipation within the poroelastic material):

\begin{equation}
    e^2={e_{ref}}^2-\alpha_{Adh}-\alpha_{Film}-\alpha_{Poro}
    \label{eq:1-alpha1-alpha2-alpha3}
\end{equation}
\subsection*{Comparison with experiments}

\begin{figure}[H]
    \centering
    \includegraphics[width=1\textwidth]{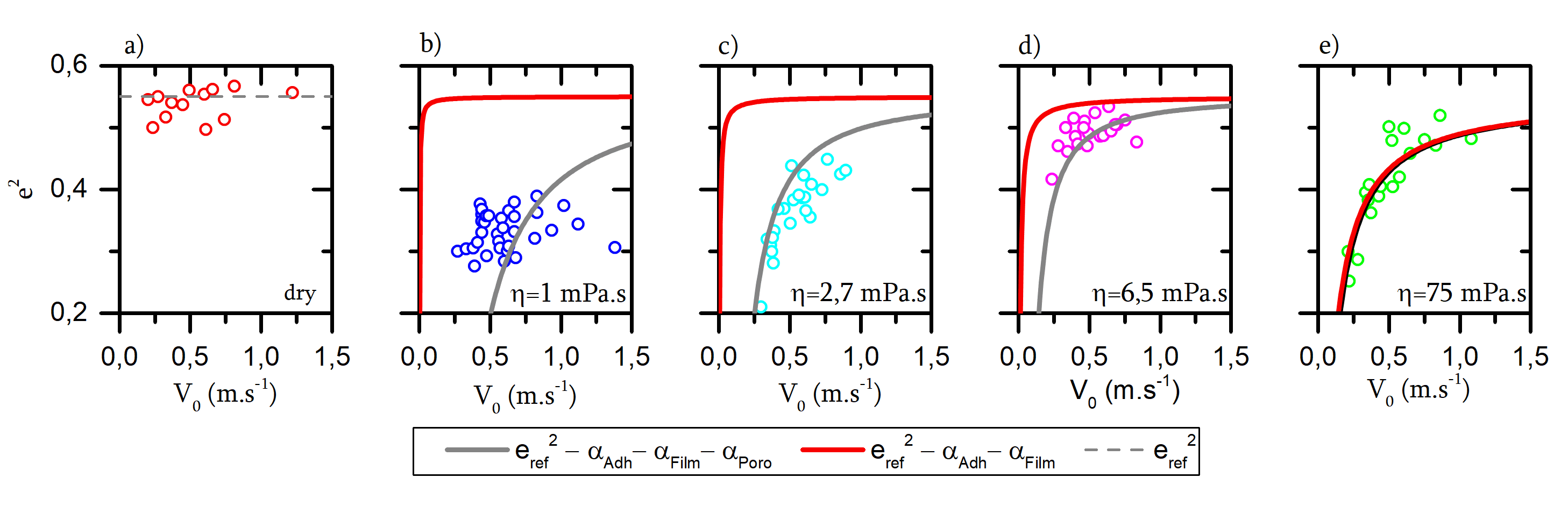}
    \caption{Measured restitution coefficients $e^2$ vs. impact speed $V_0$ for foam balls saturated with different fluids: air (a), water (b), water-glycerol mixtures (c and d) and rapeseed oil (e). The dry reference restitution coefficient is $e_{ref}^2 = 0.55$. Also shown in panels b)-e) are the predictions (red lines) using the equation \ref{eq:1-alpha1-alpha2} combining the capillary adhesion and viscous film dissipation model with
    a critical Stokes number $St_c=3.2$. Finally, the black line in each panel shows the prediction of the solution of equation (\ref{eq:1-alpha1-alpha2-alpha3}) with a fitting parameter $B=0.05$ and $St_c=3.2$. The other parameters used in the models are given in Tables \ref{Table: ball_properties} and \ref{Table:  liquids_properties} in the Methods section.}
    \label{fig:COR with models}
\end{figure}

We performed experiments with foam balls saturated with different viscous fluids. Figure \ref{fig:COR with models} shows the measured restitution coefficients $e^2$ as functions of the impact velocity $V_0$ for the different cases. 
For a dry ball, $e^2\approx 0.55$ is nearly constant (Figure \ref{fig:COR with models}a). For balls saturated with water, $e^2\approx 0.3$ is also nearly constant, but the data show considerable scatter (Figures \ref{fig:COR with models}b). For balls saturated with water-glycerol mixtures, $e^2$ increases with $V_0$, especially for water-glycerol I (Figures \ref{fig:COR with models}c-d). Finally, for a ball saturated with viscous oil $e^2$ increases strongly with $V_0$. (Figure \ref{fig:COR with models}e).

Also shown in Figure \ref{fig:COR with models} are the predictions of the viscous film dissipation model (red lines) and the solution of equation (\ref{eq:1-alpha1-alpha2-alpha3}) (gray lines).
Calculation of the porous flow model predictions requires an estimate of the permeability, which we determined to be $\kappa\approx 70 \times  10^{-12}$ m$^2$ (see Supplementary Information). Using this value, we find reasonable agreement with the observed variation of $e^2$ versus impact velocity for balls saturated with water and water-glycerol mixtures (Figures \ref{fig:COR with models} b, c and d).
However, a curious result is found for a ball saturated with rapeseed oil, whose viscosity is sufficiently high that viscous film dissipation alone can explain the measurements (Figure \ref{fig:COR with models} e). Using $St_c$ as the only adjustable parameter, we find that $St_c = 3.2$ gives the best fit for this case. Because porous flow dissipation diminishes with increasing viscosity, its contribution is too small to cause any observable effect in this case. A partial comparison between the different dissipation mechanisms is shown in the SI.

\begin{figure}[!ht]
    \centering   \includegraphics[width=0.65\textwidth]{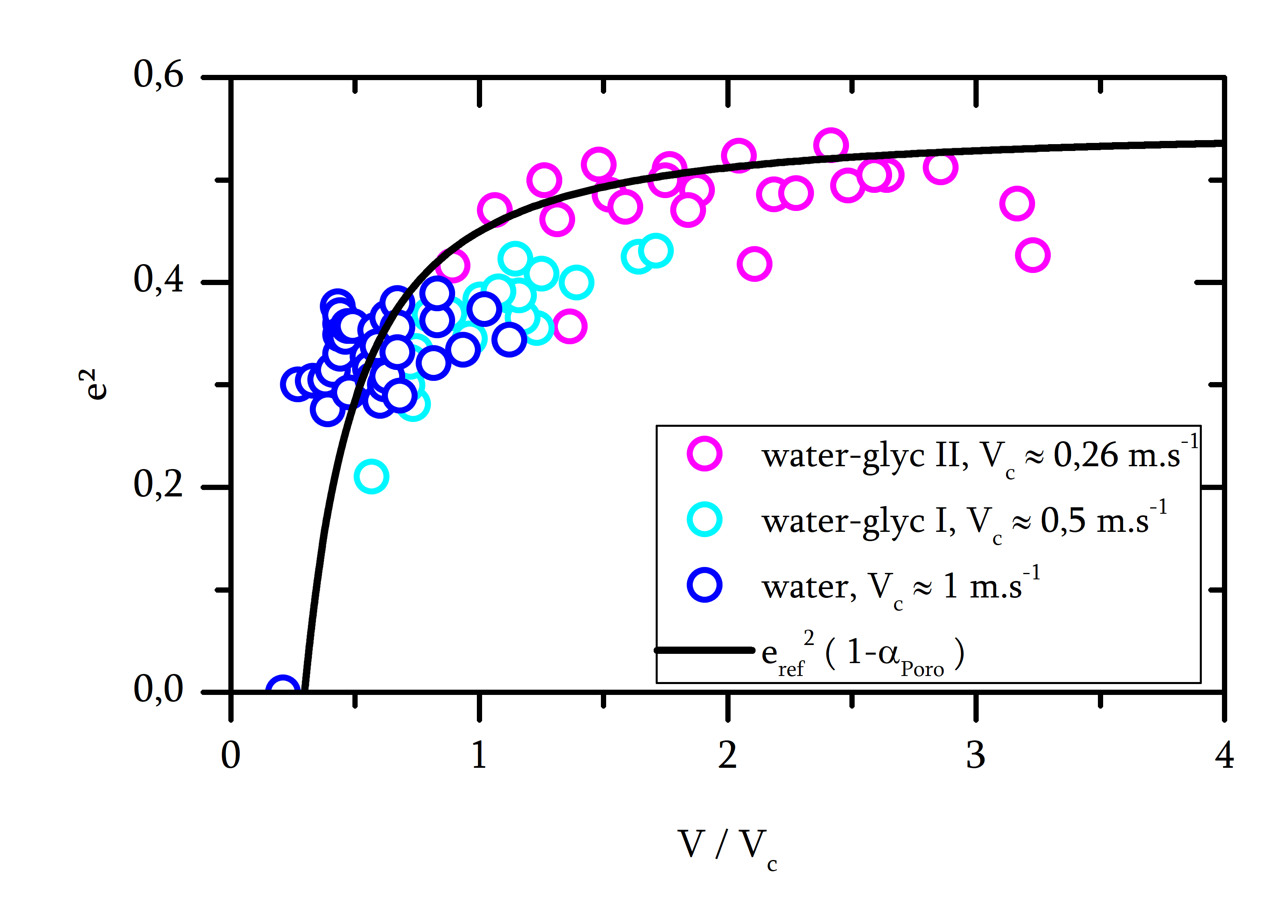}
    \caption{Coefficient of restitution of foam balls saturated with water and water-glycerol mixtures rescaled on a single curve using the critical impact velocity $V_c$ given in equation \ref{eq: Vc}.}
    \label{fig:master-curve}
\end{figure}

The dominance of porous flow dissipation for water- and water/glycerol-saturated foam balls allows us to simplify the presentation of the experimental results for these cases. Figure \ref{fig:master-curve} shows the experimentally determined restitution coefficients plotted against the ratio $V_0/V_c$, where $V_c$ is the characteristic velocity scale (see Supplementary Informations). To within error due to experimental scatter, the data collapse onto a single master curve. The agreement between our model and the experimental results for fluids of different viscosity as well as the rescaling of the data onto a master curve confirm the importance of internal poroelastic flows in dissipating energy due to impacts.


Returning to the case of hydrogel balls, we note that poroelastic dissipation is too small to explain the data. It has already been shown that the permeability of a hydrogel is on the order of $\kappa \sim 10^{-16}$ m$^2$ or smaller \cite{dasgupta_microrheology_2005,abidine_physical_2015,fujiyabu_permeation_2017}. For such low permeabilities, the contribution of viscous dissipation in the pores is negligible. For low permeability and very viscous fluids, porous balls simply act as elastic materials. 

\section*{Discussion}


We have examined the impact behavior of different soft solids on rigid surfaces, including solid rubber balls, hydrogel balls, and dry or liquid-saturated foam balls. While the deformation of such soft solids during impact is well described by Hertz contact theory, their rebound behavior can only be explained by invoking a variety of dissipation mechanisms. For each type of soft solid, we measured the restitution coefficient $e^2$ as a function of the impact velocity $V_0$. 
For the rubber and dry foam balls, $e^2$ is independent of the impact velocity but less than unity, indicating the presence of `dry' dissipation mechanisms that are beyond the scope of our work. The situation is considerably more complicated and interesting for the `wet' cases of hydrogel balls and fluid-saturated foam balls. For these cases, $e^2$ is an increasing function of $V_0$, meaning that more rapidly impacting balls rebound more efficiently. We identified three possible dissipation mechanisms that may explain this behavior: capillary adhesion, dissipation in the viscous film between the ball and the substrate, and internal poroelastic dissipation.

\begin{figure}[!ht]
    \centering
    \includegraphics[width=0.84\textwidth]{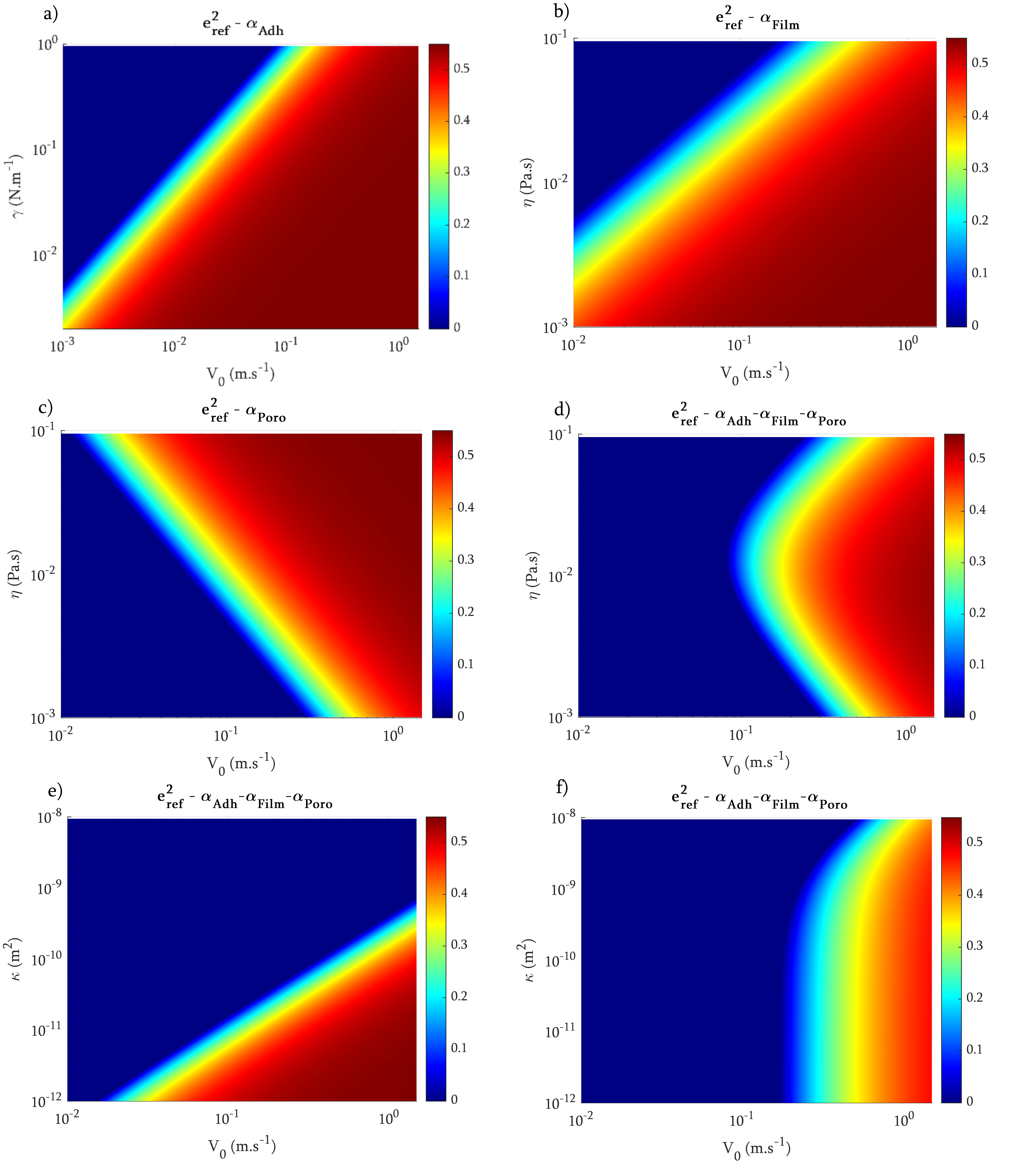}
    \caption{Maps of $e^2$ for foam balls impacting a solid surface. Variables are the impact velocity and the liquid viscosity or surface tension. a) Solution of dissipation from capillary adhesion (equation \ref{eq: COR adhesion model}). b) Solution of viscous dissipation in the liquid film (equation \ref{eq: COR EHL model}). c) Solution  of viscous dissipation within the porous sphere (equation \ref{eq: COR poroelastic model}). d) Solution of the sum of the three dissipative terms. The surface tension is taken as that of pure water $\gamma=72$mN.m$^{-1}$. The free parameter is fixed as $B=0.05$ and the critical Stokes number as $St_c=3.2$. (e-f) Solution of the sum of the three dissipative terms depending on the permeability of the foam ball and for two different viscosity $\eta=$1mPa.s (Figure e) and $\eta=$100mPa.s (Figure f). The surface tension $\gamma$, the critical Stokes number $St_c$ and the free parameter $B$ are fixed as $\gamma=72$mN.m$^{-1}$, $B=0.05$ and $St_c=3.2$.}
    \label{fig:colormap}
\end{figure}

We propose a new model for poroelastic dissipation: this model combines the Hertz model with Darcy's law and gives rise to a solution of $e^2$ which increases with the impact velocity. We found that the dissipative term is proportional to $\kappa/\eta$: a highly permeable material will let the liquid flow easily through it which leads to strong dissipation. On the other hand, a high viscosity of the fluid will allow less flow and almost no energy is dissipated. 
In contrast with fluid-saturated foam balls, the rebound systematics of hydrogel balls can best be explained by a combination of capillary adhesion and viscous film dissipation, with only a minor contribution from poroelastic dissipation. We believe that this is due to the very low permeability of hydrogel balls, which is 4-5 orders of magnitude smaller than that of the foam balls used. Because the rate of poroelastic dissipation is proportional to the permeability, this contribution to the dissipation can be expected to be small for hydrogel balls compared to the foam balls. The rate of poroelastic dissipation is also inversely proportional to the fluid viscosity, which explains why this dissipation mechanism appears to be negligible for foam balls saturated with viscous rapeseed oil. 

The importance of the three different dissipative mechanisms is illustrated on Figure \ref{fig:colormap}. We first estimate in Figure \ref{fig:colormap}a the energy loss due to capillary adhesion which is found to be effective only for very small velocities and high interfacial tensions. Figures \ref{fig:colormap}b and c show respectively the energy loss due to the presence of a viscous film and the poroelastic dissipation as a function of the viscosity of the internal fluid and the velocity. These color maps show that the effect of the viscous film can be observed strongly for the low velocity end and that this effect becomes stronger and effective at even higher velocities as the viscosity of the fluid increases. On the other hand, the dissipation due to the porosity of the material is effective for small viscosities for which the effect is strong even at higher velocities but decreases as the viscosity increases. The effect for high viscosities is effective only at the low velocity end. We show on Figure \ref{fig:colormap}d the sum of the different dissipative terms to illustrate the total energy loss for a porous wet ball impacting a dry surface. The most important contributions turn out to be porous dissipation at low viscosities and dissipation due to the presence of a viscous liquid film at high viscosities. To further illustrate the role of the different parameters in setting the restitution coefficient, we plot in Figures \ref{fig:colormap}e and f the sum of the three dissipation mechanisms in the permeability velocity space for two different viscosities. For the small viscosity, the effect is more important and is effective for a large velocity range for high values of the permeability. For the larger viscosity, the effect is important mostly for the small velocities.

\begin{figure}[!ht]
    \centering
    \includegraphics[width=1\textwidth]{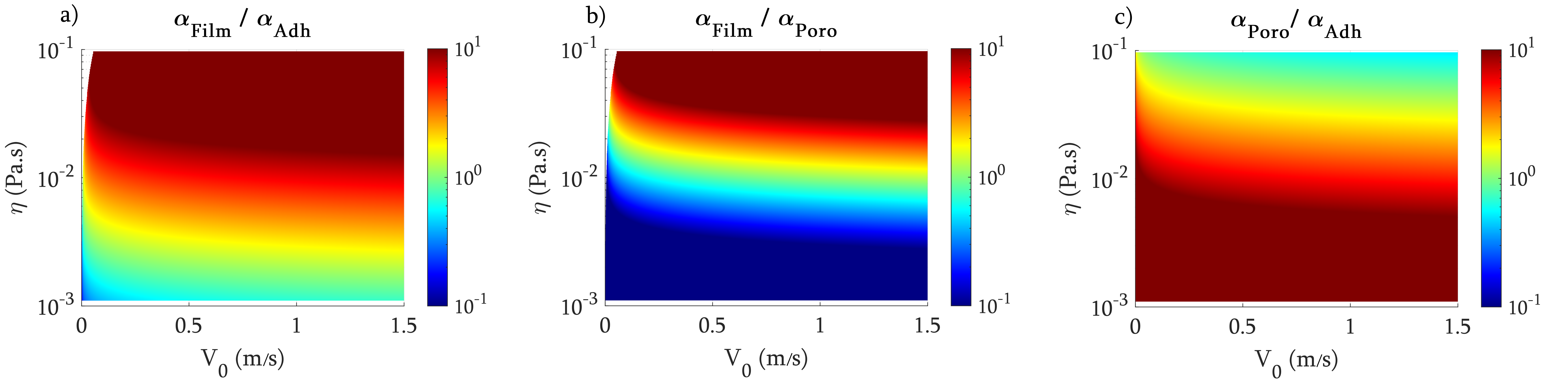}
    \caption{Phase diagrams ($V_0$, $\eta$) illustrating the different ratios of dissipative terms of a liquid-filled foam ball impacting a solid surface: a) $\frac{\alpha_{Film}}{\alpha_{Adh}}$, b) $\frac{\alpha_{Film}}{\alpha_{Poro}}$ and c)  $\frac{\alpha_{poro}}{\alpha_{Adh}}$. The two parameters are set as B=0.05 and $St_c=3.2$. Surface tension is fixed as $\gamma = 72 $mN.m$^{-1}$ and physical properties of the porous foam ball are given in Table \ref{Table: ball_properties}.}
    \label{fig:ratio alpha}
\end{figure}

To better illustrate the relative importance of each dissipative term, we show on Figure \ref{fig:ratio alpha} colormaps of different ratios of the different dissipative terms.
Figure \ref{fig:ratio alpha}a shows that dissipation in the viscous film becomes dominant with respect to the adhesion as the viscosity increases. Figure \ref{fig:ratio alpha}b shows that poroelasticity dominates at the lower viscosity end with respect to the film dissipation which takes over only at high enough viscosities. The poroelastic dissipation is however generally larger than dissipation due to capillary adhesion except at sufficiently high viscosities.

Our model is, however, only valid for small deformations and cannot explain the decrease of $e^2$ with increasing $V_0$ for hydrogel balls when $V_0 > 1$ m s$^{-1}$
(Figure \ref{fig:bouncing-overview}e and Figure \ref{fig:COR highspeed}).

Overall, this study provides insights into the complexity of impacts of liquid-filled soft solids, highlighting the importance of the liquid and - most importantly - the presence of liquid within the porous structure of the solids in dissipating impact energy. This last dissipation mechanism and the closed theoretical expression for its effect on the restitution coefficient open the way to designing and engineering a new generation of shock absorbers, cushions and armor.
An example of such an application is Dawson et al.'s \cite{dawson09japplmech} study of an open-cell foam impregnated with a shear-thickening non-Newtonian fluid.

\section*{Methods}

\subsection*{Experimental measurements of the foam and hydrogel balls}

The spheres used in the experiments are made of either rubber, hydrogel or foam (see Figure \ref{fig:experimental_setup}b).
The rubber balls were bought from Intertoys and are made of a synthetic rubber. 
Hydrogel balls are originally small dry beads of polymer (polyacrylamide) bought from Educational innovations (GM 710). These dry beads are immersed in a solution of salty water (pure Milli-Q Water with a concentration of $0.6 \ g/L$ of NaCl) and left to swell for $24$ hours. During this time, they reach an average diameter of $D_0\approx 15$ mm with a liquid volume fraction of $\phi=0.98$ and a constant mass of approximately $m=2.8$ g. 
The foam balls used are commercial tennis foam balls purchased from Intersport (Ref: Pro Touch 412174 6TP) and are made of polyurethane.
The foam balls have diameter $D_0 = 68$ mm and exhibit sponge-like behavior : Water or other liquids can be absorbed and released under pressure.

The mechanical response of the rubber, hydrogel and foam balls were measured using a rheometer in compression mode.

All rubber, foam and hydrogel ball properties are reported in Table S\ref{Table: ball_properties}.
The properties of the liquids used to saturate the foam balls in the experiments are given in Table S\ref{Table:  liquids_properties}. 

Before each experiment, we measure the mass of the saturated ball. The mass of the filled foam ball before each experiment is kept constant at approximately $m\approx 122 \pm 2$ g.  Once  the mass is determined, the ball is dropped by hand from a certain height and bounces multiple times. The experimental setup is shown in Figure \ref{fig:experimental_setup}.

When the hydrogel ball is removed from the water, it is fully covered with a thin film of water. Before being dropped, the excess of water is removed with a clean paper tissue. However, due to the porous nature of the gel, a thin film of water remains on the surface of the ball. The hydrogel ball is then dropped by hand from different heights. Two high-speed cameras (Phantom V641) are used simultaneously to capture the impact from the bottom and the side. The surface used is a 1.5 cm thick transparent poly-carbonate plate. This plate was solidly fixed on top of a granite table leaving a space between the table and the plate where a mirror was inserted at 45° to visualize the impact from below. Camera frame rates range from 2000 to 10000 fps for the side views and from 70000 to 80000 fps for the bottom view. After each impact, the hydrogel ball is immediately put back into water to prevent evaporation.

The foam ball is first entirely immersed by hand in a pool of liquid and compressed manually to expel the air trapped inside.
Different liquids are used to fill the ball: water, water-glycerol mixtures and rapeseed oil. 
After the pressure is released, liquid is sucked into the ball. Once the ball is back to its spherical form, it is removed from the liquid pool and it is cleaned using tissues to remove the excess of liquid present on its surface.

\section*{Acknowledgements}

We thank Vincent Bertin for discussions.

\section*{Author contributions statement} 

 B.G. carried out most of the experiments with help from H.K. and D.B., N.R. elaborated the poro-elastic model with help from all authors. The data analysis was carried out by B.G. with help from H.K. and D.B.. All authors contributed to the writing of the paper.


\bibliography{main_Ncomms_article}
\bibliographystyle{unsrt}

\newpage

\title{Impacts of poroelastic spheres - Additional informations}

\maketitle
\setcounter{figure}{0}
\makeatletter 
\renewcommand{\thefigure}{S\@arabic\c@figure}
\makeatother

\subsection*{Hertzian contact of hydrogel and dry foam balls}

The force applied on an elastic sphere in contact with a rigid surface $F_{Hertz}$ depends  nonlinearly on the total deformation $h$ according to Hertz Theory\cite{hertz_ueber_1882}:

\begin{equation}
    F_{Hertz}=\frac{4}{3}\frac{E_0}{1-\nu^2}\sqrt{R}h^{3/2}
    \label{eq: Hertz_Force}
\end{equation}

In equation \ref{eq: Hertz_Force}, $E_0$, $\nu$ and $R$ are respectively Young's modulus, Poisson's ratio and the radius of the sphere. 
During the collision, the initial kinetic energy $\frac{1}{2}m{V_0}^2$ is converted into both kinetic and elastic energy associated with the deformation.
Based on Hertz theory, Landau and Lifshitz \cite{landau_theory_1986} wrote the following equation of conservation of energy:

\begin{equation}
\frac{1}{2} k h^{5/2} + \frac{1}{2} m \Dot{h}^2 = \frac{1}{2} m {V_0}^2
    \label{eq: energy conservation hertz}
\end{equation}
Here, $k = 16\sqrt{R} E_0/(15 \left( 1-\nu^2\right))$ with $m$ and $V_0$ being respectively the mass and the impact velocity of the elastic ball.

From equation \ref{eq: energy conservation hertz} one can calculate the maximum contact diameter $D_{max}$ and the contact time $t_{max}$ required to go from $D=0$ to $D=D_{max}$:

\begin{align}
    D_{max} &=2\sqrt{R\left(\frac{m}{k}\right)^{2/5}V_0^{4/5}} &
   t_{max} &=1.47\left(\frac{m^2}{k^2V_0}\right)^{1/5}
    \label{eq: Dmax and tm Hertz}
 \end{align}

\begin{figure}[!ht]
    \centering    \includegraphics[width=1\textwidth]{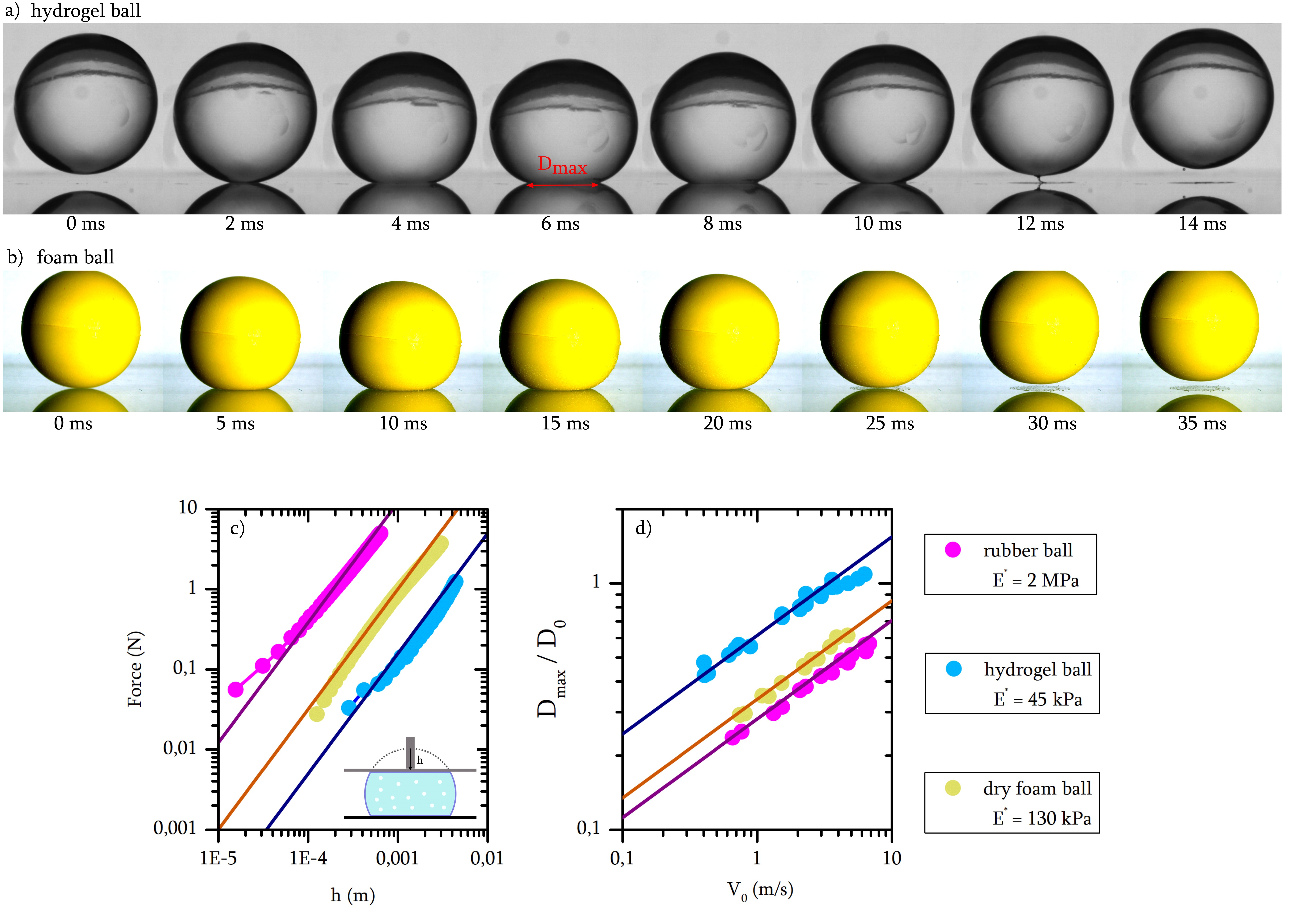}
    \caption{ a) Deformation of a hydrogel ball of radius $R=7$ mm impacting at $V_0=0.62$ m.s$^{-1}$. b) Deformation of a water filled foam ball of radius $R=34$ mm impacting at $V_0=0.47$ m.s$^{-1}$. c) Loading curves (force vs displacement $h$) of the rubber ball, the hydrogel ball and the dry foam ball. Experimental points are obtained with a rheometer in compression mode and each point corresponds to the measured force averaged over one second. Solutions of equation \ref{eq: Hertz_Force} are represented in continuous lines. d) Normalized maximum contact diameter of rubber, hydrogel and dry foam balls for different impact velocities. Continuous lines represent the solutions of the Hertz model (see equation \ref{eq: Dmax and tm Hertz}). The elastic modulus $E^*$  given in the caption is used to fit both the force vs. loading curves and the normalized maximum contact diameter.}
    \label{fig:tensile-stress}
\end{figure}

The results of the loading part are shown in Figure \ref{fig:tensile-stress}. 
The elastic modulus $E^*$ is found from the small deformation regime which follows the Hertz prediction $F\propto h^{3/2}$. 
\newpage

\newpage
\subsection*{Measurement of the permeability}

\begin{figure}[ht!]
    \centering
    \includegraphics[width=0.15\textwidth]{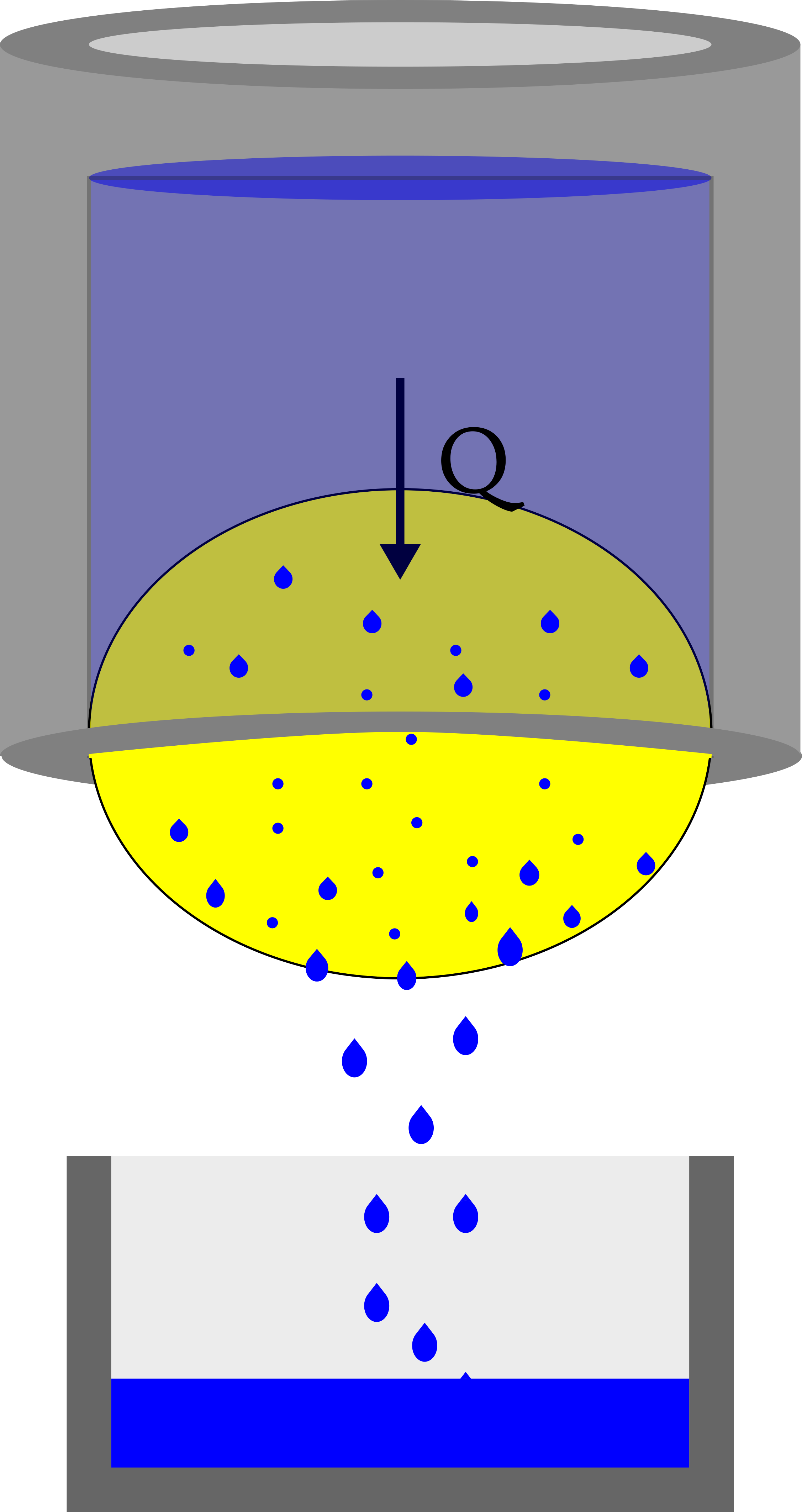}
    \caption{ Sketch of Darcy flow experiment with the foam ball.}
    \label{fig6: permeability}
\end{figure}

The permeability of the foam balls is deduced experimentally using the experimental setup shown in Figure \ref{fig6: permeability}. 

A transparent cylinder is placed vertically with a foam ball glued to its end. The cylinder is then filled with either water or oil. Due to hydrostatic pressure, the liquid flows through the foam ball. A scale with a beaker is placed beneath the tube to collect the liquid and measure the flow rate $Q = U \pi R^2$ (in m$^3$ s$^{-1}$). A camera is aligned with the cylinder to record the water height $H$ inside the tube in order to determine the hydrostatic pressure gradient $\nabla P$ as a function of time. The pressure gradient is related to the difference of pressure between the upper and lower surface of the sphere. We approximate the sphere as a cylinder of radius $R=34$ mm and of length $L=68$ mm so the pressure gradient is:

\begin{equation}
    \nabla P = \frac{P_{hydro}}{L}
\end{equation}

with the hydrostatic pressure $P_{hydro}$ defined as: 

\begin{equation}
    P_{hydro}=\rho g \pi R^2 H(t)
\end{equation}

The flow velocity $U$ is related to the flow rate $Q$ and the pressure gradient (Darcy's law)  with the following relation:

\begin{equation}
    U =\frac{Q}{\pi R^2}=\frac{\kappa}{\eta}\nabla P
\end{equation}

\begin{figure}[!ht]
    \centering
    \includegraphics[width=0.8\textwidth]{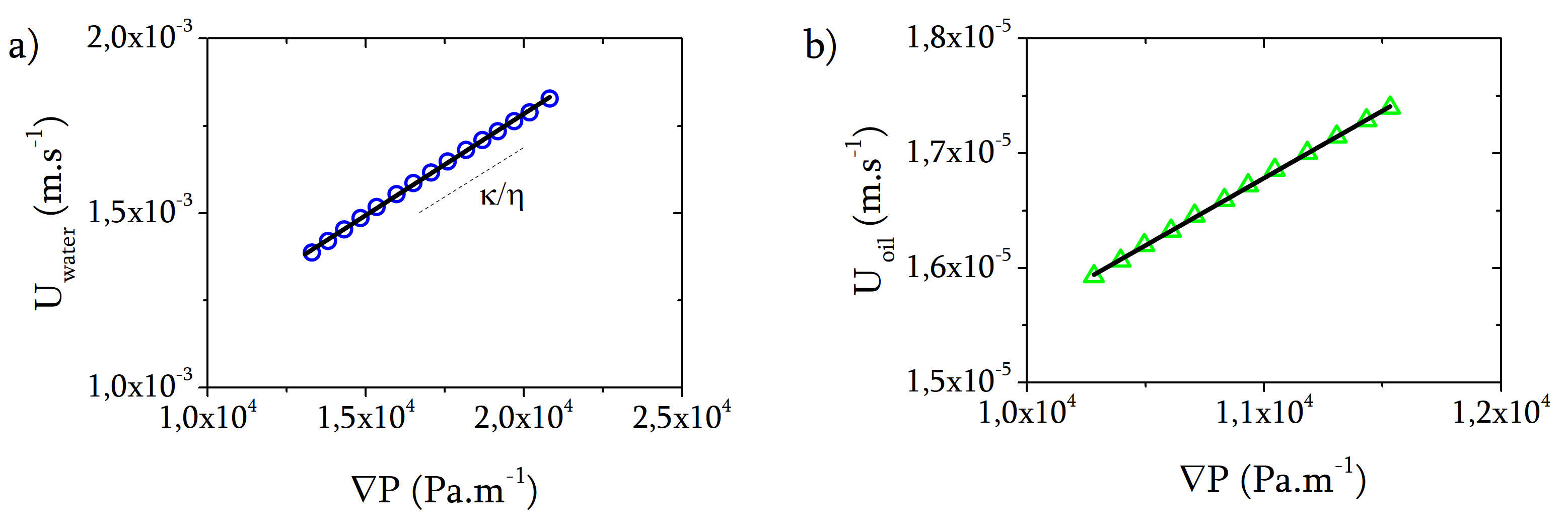}
    \caption{ Measurement of the permeability of the foam ball with two different liquids : a) water ($\eta= 1$ mPa.s) and b) rapeseed oil ($\eta = 75$ mPa.s)}
    \label{fig:darcy law}
\end{figure}

Knowing the viscosity of the fluid (see table \ref{Table:  liquids_properties}), the permeability $\kappa$ can be determined by placing the ball in the column filled with a liquid and by using equation 7 in main text. The flow velocity $U$ follows a linear trend with the pressure gradient for both water and oil as reported in Figure \ref{fig:darcy law}. The permeability is given by the slope of each curve, yielding an average permeability $\kappa \approx 74 \mu m^2$.

\subsection*{Model of energy dissipation due to capillary adhesion}

\begin{figure}[ht!]
    \centering
    \includegraphics[width=0.7\textwidth]{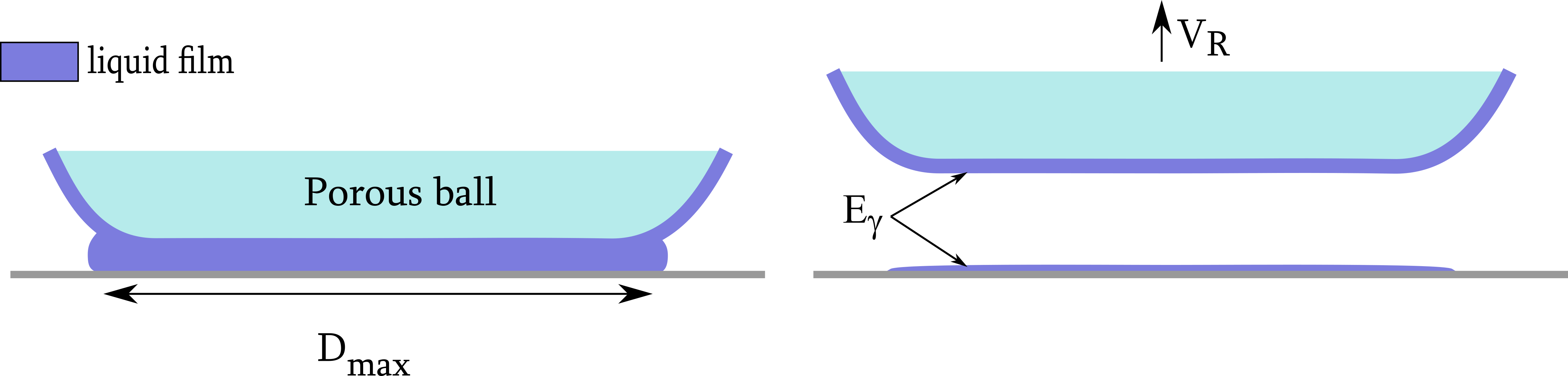}
    \caption{ model of energy dissipated from capillary adhesion.}
    \label{fig5: adhesion energy model}
\end{figure}

Due to the porous nature of the hydrogel balls or the sponge balls filled with fluid, a liquid film is always present between the sphere and the surface. It costs energy to detach the sphere from the surface as two new liquid-air interfaces need to be created (see figure \ref{fig5: adhesion energy model}).  

In this case, the conservation of energy needs to take into account an additional contribution from the capillary forces. We have:

\begin{equation}
    \frac{1}{2}m{V_r}^2+E_{\gamma}=\frac{1}{2}m{V_0}^2
    \label{eq: nrj_cons_adhesion}
\end{equation}

with $V_0$ the impact velocity, $V_r$ the rebound velocity, $m$ the mass of the ball and $E_{\gamma}$ the capillary adhesion energy. We assume that the ball is perfectly elastic so the maximum contact radius $R_{max}$ is given by the Hertz model (see equation \ref{eq: Dmax and tm Hertz}) and the capillary adhesion energy can be written as:

\begin{equation}
    E_{\gamma}=\frac{1}{2} \gamma \pi {D_{max}}^2 
\end{equation}

By using equation \ref{eq: Dmax and tm Hertz}, we have, for $e^2$ as a function of impact velocity $V_0$:

\begin{equation}
      e^2 = 1-\frac{4\gamma \pi R\left(\frac{m}{k}\right)^{2/5}}{m}V_0^{-6/5}\equiv 1 - \alpha_{Adh}.
      \label{eq: COR adhesion model}
\end{equation}

Note that $e^2$ increases as the impact velocity increases, which is consistent with our observations. However, the limit of equation \ref{eq: COR adhesion model} when $V_0$ goes to $0$ is not realistic. In reality, below a critical velocity $V_{crit}$, the ball no longer bounces and instead sticks to the surface resulting in $e^2 = 0$. 


\subsection*{Model of viscous dissipation in the lubricated liquid film}

\begin{figure}[ht!]
    \centering
    \includegraphics[width=0.4\textwidth]{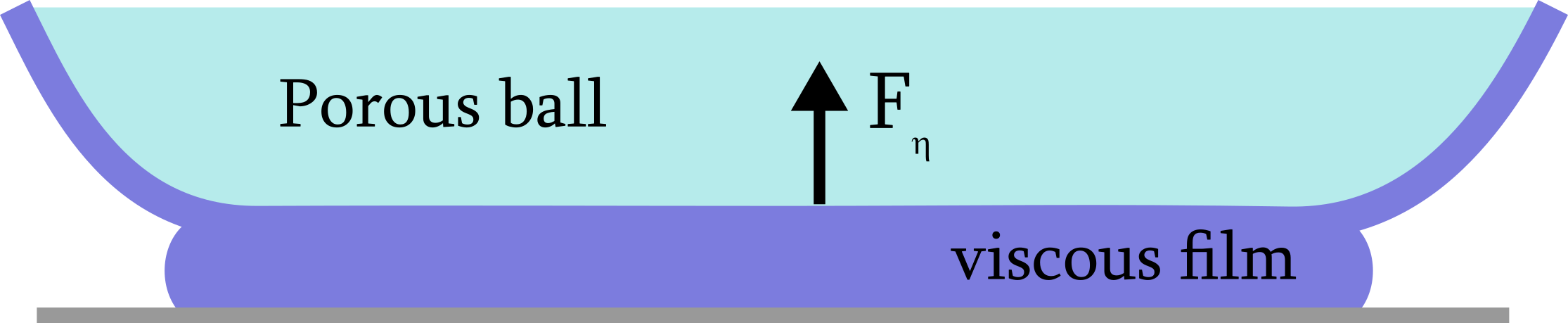}
    \caption{ Model of viscous dissipation due to the presence of a viscous film between the elastic sphere and the surface.}
    \label{fig5: EHL model}
\end{figure}

\begin{figure}[!ht]
    \centering
    \includegraphics[width=1\textwidth]{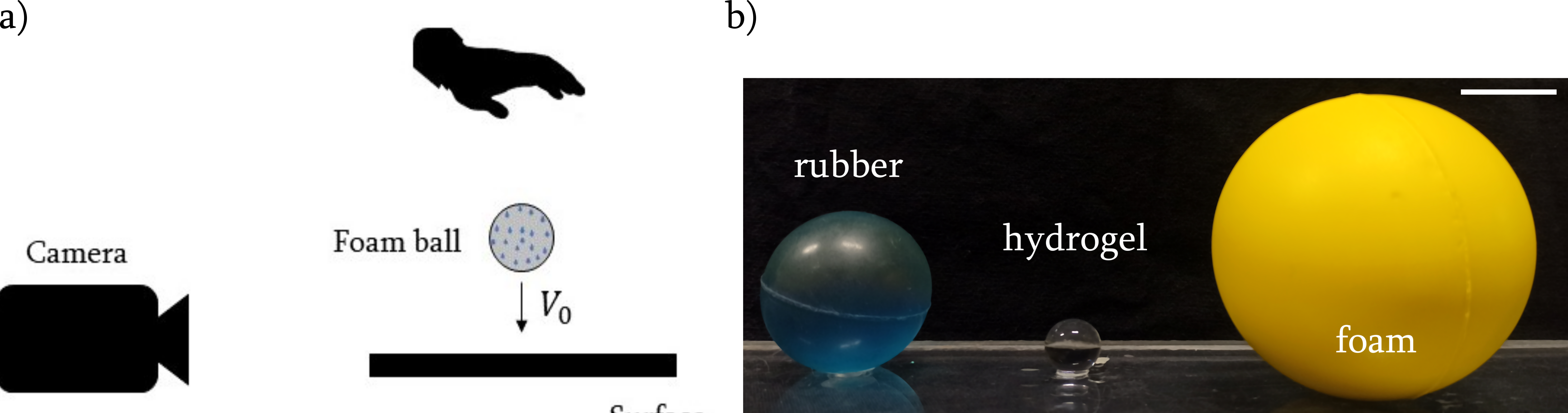}
    \caption{ a) Experimental setup for hydrogel and foam ball impacts. b) Size comparison between the rubber ball (left), the hydrogel ball (center) and the foam ball (right). Scale bar: 2cm}
    \label{fig:experimental_setup}
\end{figure}

\begin{figure}[ht!]
    \centering
    \includegraphics[width=0.85\textwidth]{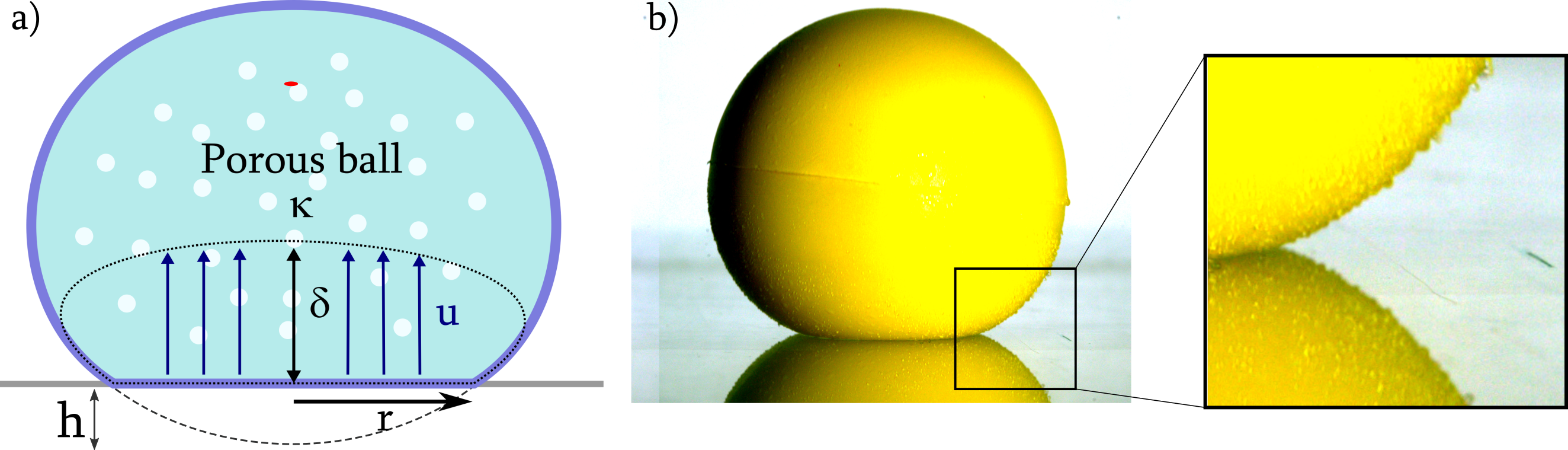}
    \caption{ a) Representation of dissipation from the flow inside the porous network of the hydrogel induced by the Hertz pressure during the impact. b) Picture of the water filled foam ball impacting at $V_0=0.47 \ m.s^{-1}$  and where small water droplets are expelled from the pores during the impact.}
    \label{fig5: poroelastic model}
\end{figure}

\begin{figure}[ht!]
    \centering
    \includegraphics[width=\textwidth]{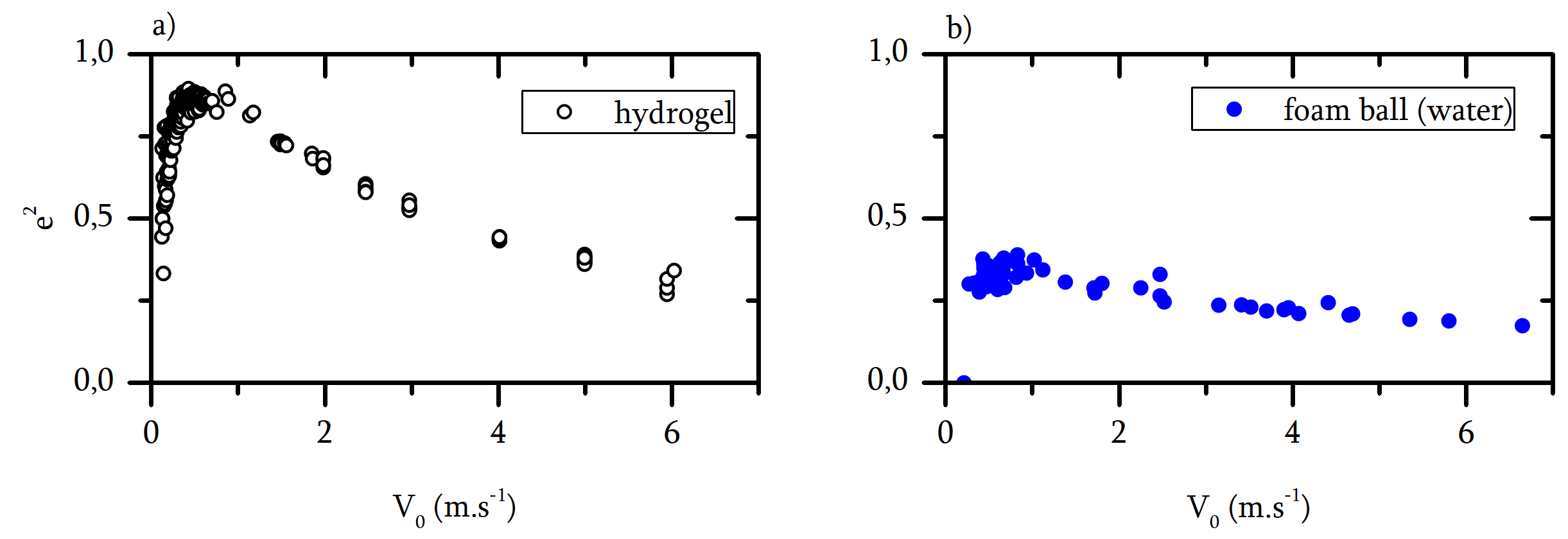}
    \caption{ Coefficient of restitution of a) hydrogel and b) water-filled foam balls for different impact velocities. }
    \label{fig:COR highspeed}
\end{figure}

The second dissipation mechanism we consider comes from the presence of a liquid film underneath the ball. 
During the impact, the liquid film squeezed between the porous ball and the surface acts as a lubrication layer and affects the contact dynamics. When a sphere impacts this wet surface, a portion of the initial kinetic energy is lost through viscous dissipation in the liquid film \cite{davis_elastohydrodynamic_1986,barnocky_elastohydrodynamic_1988,gondret_experiments_1999,gondret_bouncing_2002,davis_elastohydrodynamic_2002} (see Figure \ref{fig5: EHL model}).

Davis et.al \cite{davis_elastohydrodynamic_2002} proposed a simple model to explain the collision of a ball with a surface coated with a viscous film. The viscous lubrication force resisting the approach of the ball toward the surface is:

\begin{equation}
    F_L= \frac{6 \pi \eta R^2 v}{x}
\end{equation}

where $x$ is the minimum distance between the surface of the sphere and the substrate and $v=-\mathrm dx/\mathrm dt$ is the relative velocity of the sphere. By writing:
\begin{equation}
    m \frac{dv}{dt} = -F_L
    \label{eq: Davis kinetic}
\end{equation}
and integrating equation \ref{eq: Davis kinetic} with the initial conditions $v_{t=0}=V_0$ and $x_{t=0}=x_0$ (the initial thickness of the viscous liquid film),
they found the following relation for the coefficient of restitution: 
\begin{equation}
    e^2 = \left( 1-\frac{St_c}{St}\right)^2 \equiv 1 - \alpha_{Film}.
    \label{eq: COR EHL model}
\end{equation}

Here $St=m V_0/(6 \pi \eta R^2)$ is the Stokes number, $St_c=\ln(x_0/x_r)$ is the critical Stokes number at which the ball no longer bounces, and $x_r$ the relative distance between the sphere and the surface when the ball bounces back. Considering that the thickness of the liquid film is between 10$\mu$m to 1mm and that $x_r$ can not be smaller than the highest roughness value between the impacted surface and the ball \cite{davis_elastohydrodynamic_1986,davis_elastohydrodynamic_2002}. Roughness of plexiglas surface is of the order of $0.1-1 \mu m$ \cite{range_influence_1998} while roughness of the hydrogel ball is estimated as the pore size which is of the order of 10 nanometers \cite{fujiyabu_permeation_2017}. Pore size of the foam ball is of the order of 1 to 100 $\mu m$. Therefore, $St_c$ for hydrogels are estimated from the plexiglas roughness and taken between $6<St_c<10$ and realistic values of $St_c$ for foam balls are $0<St_c<7$.
As for adhesion,  equation \ref{eq: COR EHL model} shows an increase of $e^2$ as $V_0$ increases.
In this model, the critical Stokes number $St_c$ is difficult to estimate and is used as a free parameter constrained by the condition that below $St_c$ the rebound no longer occurs \cite{davis_elastohydrodynamic_2002}. 

\section*{Experimental setup}

\newpage

\section*{Coefficient of restitution of hydrogel and water filled foam balls}

\section*{Liquid and solid properties}

\begin{table}[!ht]
    \centering
    \begin{tabular}{||c c c c c||}
    \hline
     material & $E^*$ (kPa) & $\kappa$ (m$^{2}$) & m (g) & $R$ (mm) \\
     \hline\hline
     hydrogel ball & $45$ & $10^{-15}$ - $10^{-18}$ & $2,8$ & $\approx$ $15$ \\
     \hline
     foam ball & $130$ & $74.10^{-12}$ & $122$ & 68 \\
     \hline
     rubber ball & $2000$ & -- & $34.6$ & $21$
     \\
     \hline
    \end{tabular}
    \caption{$E^*=\frac{E}{1-\nu^2}$ with $E$ the elastic modulus and $\nu$ the poisson ratio. $\kappa$ is the permeability \cite{dasgupta_microrheology_2005,abidine_physical_2015}, $m$ the mass  and $R$ the radius of hydrogel, foam and rubber balls.}
    \label{Table: ball_properties}
\end{table}

\begin{table}[!ht]
    \begin{center}
    \begin{tabular}{||c c c c||} 
     \hline
     liquid & $\rho$ ( kg m$^{-3}$) & $\eta$ (mPa s) & $\gamma_{LV}$ ( mN m$^{-1}$) \\ [0.5ex] 
     \hline\hline
     water & 1000 & 1 & 72 \\ 
     \hline
      Glycerol-Water 1 & 1031 & 2,7 &  64 \\
     \hline
      Glycerol-Water 2 & 1138 & 6,5 &  64 \\
     \hline
     Rapeseed oil & 920 & 75 &  33 \\
     \hline
     \end{tabular}
    \end{center}
    \caption{Liquids properties used in the experiments. $\rho$ is the liquid density, $\eta$ is the dynamic viscosity and $\gamma_{LV}$ is the surface tension of the liquid with air.} 
    \label{Table:  liquids_properties}
\end{table}

\end{document}